\documentclass{amsart}

\usepackage{amsmath}
\usepackage{amsthm}
\usepackage{amssymb}
\usepackage{graphicx}
\usepackage{hyperref}
\usepackage{amsaddr}

\newcommand{\G}{GV}
\newcommand{\Hh}{\mathcal{H}}

\title[Godbillon-Vey in Ideal Fluids]{The Godbillon-Vey Invariant as Topological Vorticity Compression and Obstruction to Steady Flow in Ideal Fluids}
\author{Thomas Machon}
\address{H.H.~Wills Physics Laboratory, Tyndall Avenue, Bristol BS8 1TL, UK}
\begin{document}

\maketitle

\begin{abstract}
If the vorticity field of an ideal fluid is tangent to a foliation, additional conservation laws arise. For a class of zero-helicity vorticity fields the Godbillon-Vey (GV) invariant of foliations is defined and is shown to be an invariant purely of the vorticity, becoming a higher-order helicity-type invariant of the flow. $\G \neq 0$ gives both a global topological obstruction to steady flow and, in a particular form, a local obstruction. $\G$ is interpreted as helical compression and stretching of vortex lines. Examples are given where the value of $\G$ is determined by a set of distinguished closed vortex lines.
\end{abstract}

\section{Introduction}

In an ideal fluid the evolution of vorticity is given by
\begin{equation}
\partial_t \omega+ \mathcal{L}_U \omega=0,
\label{eq:vort}
\end{equation}
where $U$ is the fluid velocity, $\omega$ the vorticity and $\mathcal{L}$ the Lie derivative. In a domain $\Omega$, the fluid velocity is a vector field satisfying the conditions
\begin{equation}
 \nabla \times U = \omega, \quad \nabla \cdot U =0,\quad U \parallel\partial\Omega,
\end{equation}
which determine $U$ up to the addition of a harmonic vector field. 
\eqref{eq:vort} implies that vortex lines flow along $U$, they are `frozen' in the field. Indeed, if $\Phi^t$ is the diffeomorphism of $\Omega$ generated by the fluid after a time $t$, then the vorticity satisfies
\begin{equation}
\omega_t = d\Phi^t (\omega_0).
\label{eq:trans}
\end{equation}
This transport of vorticity leads to a number of conservation laws, in particular, as first observed by Woltjer~\cite{woltjer58} and Moreau~\cite{moreau61}, the helicity of the vorticity field
\begin{equation}
\Hh = \int_\Omega U \cdot \omega \;dV,
\end{equation}
is conserved by the fluid. This can be seen explicitly, 
\begin{equation}
\frac{d \Hh}{d t} = \int_\Omega (\partial_t + U \cdot \nabla)( U \cdot \omega )  dV = \int_\Omega \nabla \cdot (F \omega) dV,
\end{equation}
where $F= U^2/2-P$, and $P$ is pressure. Stokes' theorem then gives a boundary term which vanishes as long as $\omega\parallel\partial \Omega$, a condition we impose henceforth. $\Hh$ is a topological quantity invariant not only under the flow of $U$ but any divergence-free vector field. First understood by Moffatt~\cite{moffatt69}, formalised by Arnold~\cite{arnold74}, this topological aspect of helicity manifests in its relation to the average linking of vortex lines, a fact revealed by writing $U$ in terms of the Biot-Savart operator.

For curves in three dimensions, linking is perhaps the most elementary topological property and integral invariants that measure other topological aspects of $\omega$ have long been sought\footnote{Arnold and Khesin's ``dream''~\cite{arnold99}.}, with qualified success. For vorticity fields supported on handlebodies, one can define invariants that measure higher-order topological invariants (such as triple-linking numbers)~\cite{berger90, komendarczyk09, komendarczyk10, akhmetiev05,akhmetev09,deturck13,deturck13b,laurence00}, typically mirroring their development by Milnor through the use of Massey products~\cite{milnor54}, however these invariants have not been extended to general vorticity fields. Notably it has been shown by Enciso, Peralta-Salas, and Torres de Lizaur~\cite{enciso16}, building on the work of Kudryavtseva~\cite{kudryavtseva14,kudryavtseva16}, that any such invariant defined by a functional whose derivative can be written as an integral operator with a continuous kernel is a function of helicity. A manifestation of this result can be seen in the attempts to define asymptotic invariants other than helicity~\cite{gambaudo01,baader12,komendarczyk14}, which become functions of helicity.

As noted~\cite{enciso16}, this constraint can be evaded by considering invariants which are not continuous functionals on the space of volume-preserving vector fields, as in the case of the triple-linking invariants. Here we consider vorticity fields whose integral curves lie tangent to a (singular) codimension-1 foliation $\mathcal{F}$ of $\Omega$. This property is preserved under volume-preserving diffeomorphisms so that diffeomorphism invariants of $\mathcal{F}$ become invariants of the vorticity field. In particular we study vorticity fields admitting a vector potential $A$ satisfying $A \cdot \omega =0$. By Frobenius' integrability theorem, this implies that $A$ defines a codimension-1 foliation of the fluid domain. It also implies that helicity vanishes, when it is defined. In such cases we may define a higher order integral invariant, $GV$, related to the Godbillon-Vey invariant of foliations.

In this paper we discuss $GV$ for ideal fluid flows. We show, under assumptions on the singularities in $\mathcal{F}$, that it is defined purely in terms of the vorticity field, so is an integral invariant (conserved quantity) of the vorticity under general volume-preserving diffeomorphisms and hence the fluid flow. If $\Omega$ is closed and the foliation $
\mathcal{F}$ is non-singular, then $GV$ is equal to the Godbillon-Vey invariant of $\mathcal{F}$. In general, $GV$ may be defined for fluids on manifolds with boundary with $\omega \parallel \partial \Omega$, where the Godbillon-Vey invariant cannot be defined. 

As we will discuss, $GV$ is a topological invariant measuring the helical compression of the vortex lines. Insomuch as it is a combination of twisting and squeezing of vorticity, we are tempted to refer to it as the `wring' of a vorticity field, as in `wringing a towel'. While a local geometric meaning for $GV$ can be given fairly easily, the global topological interpretation is subtle~\cite{hurder02}. Here we show, for a class of vorticity fields with $\G \neq 0$, that $\G$ is determined by the local structure of a distinguished set of closed vortex lines, associated to the Kupka phenomenon of integrable 1-forms~\cite{kupka64,camacho78,camacho82}. These distinguished closed lines can be given local self-linking terms and $\G$ measures, in part, the failure of these self-linking terms to balance the linking terms from the other lines.

Helicity gives a lower bound for the enstrophy~\cite{arnold99}, and the goal has often been to give further bounds for the energy for higher-order helicities. $GV$ is dimensionless, and so cannot bound a dimensionful quantity such as the energy alone. We show, however, that $\G$ has a very natural relationship to Euler dynamics. In particular $\G \neq 0$ implies that the flow is not steady and
\begin{equation}
\G^2 \leq C \int_\Omega (\partial_t \omega)^2 dV,
\label{eq:bound}
\end{equation}
where $C$ is a positive function depending on $\omega$ and the metric of $\Omega$ for which we give an explicit form. As $\G$ is invariant under volume-preserving diffeomorphisms, this gives a global topological obstruction to the existence of a metric making $\omega$ a steady solution of the Euler equations. We show further that in a particular metric-dependent form the density for $\G$ gives a local obstruction to $\partial_t \omega$ vanishing. In this form \eqref{eq:bound} can be further approximated, giving the relation
\begin{equation}
|\G|\lessapprox \frac{L^7}{4 E^2}  \int_\Omega (\partial_t \omega)^2 \;dV,
\end{equation}
where $L$ is a characteristic lengthscale of the system as $E$ is the total kinetic energy (of $U$). In this way, $GV$ can be seen to measure the ratio between the rate of change of vorticity to the kinetic energy of the flow. As discussed above, $\G$ has a geometric interpretation, measuring the average helical vortex compression (`wring') of the flow. Vorticity compression is the core nonlinearity of the Euler equations, given this as well as \eqref{eq:bound} we speculate that flows with $\G \neq 0$ will prove particularly interesting from a dynamical perspective.

The relevance of $GV$ in hydrodynamics has been noted previously~\cite{arnold99,webb14, taba90}. In particular, Tur and Yanovsky~\cite{tur93}, establish the local conservation of $\G$, where it arises in the case of a hydrodynamic system described by a 1-form $S$, evolving as $(\partial_t +L_U) S =0$. In Webb et al.~\cite{webb19} the potential application of $GV$ to ideal fluids is discussed. In general, the vector potential $A$ for $\omega$ satisfying $A \cdot \omega =0$ may be written as $A = U + V$, where $U$ is the fluid velocity and $V$ is curl-free. Webb et al.~discuss the special case $V=0$, however, as we discuss below, $GV$ can be defined in more generality. Because $\G$ is invariant under all volume-preserving diffeomorphisms, it appears as a Casimir-type invariant in the Hamiltonian formulation of ideal fluids~\cite{morrison98}, which we discuss in another paper~\cite{machon20}.

\section{The Godbillon-Vey Invariant of a Vorticity Field}
\label{section:2}

We consider a vorticity field $\omega$ defined on a 3-manifold $\Omega$ with volume-form $\mu$.  We take all vector fields and differential forms to be smooth throughout. We impose $\omega \parallel\partial \Omega$ and
\begin{equation}
\textrm{div}(\omega) = \mathcal{L}_\omega \mu = 0,
\end{equation}
hence (by Cartan's magic formula) $\iota_\omega \mu$ is closed, assumed exact so $\iota_\omega \mu =d \theta$. Then the helicity
\begin{equation}
\Hh = \int_\Omega \theta \wedge d \theta,
\end{equation}
is an invariant of volume-preserving diffeomorphisms. Helicity is invariant under a gauge transformation, $\theta \to \theta + d f$, for any function $f$ and is invariant for {\em any} choice of $\theta$ only if it satisfies the fluxless condition~\cite{cantarella10} (see Appendix I). 
This condition is equivalent to the statement that
\begin{equation}
\int_{S} d \theta =0,
\end{equation}
for any surface $S \subset \Omega$, with $\partial S \subset \partial \Omega$. The fluxless condition is always satisfied if each component of $\partial \Omega$ is simply connected. 

Here we will consider the special case where $\theta$ can be chosen such that 
\begin{equation}
\theta \wedge d \theta = 0.
\label{eq:int}
\end{equation} 
{\it i.e.} the helicity density vanishes. In the fluxless case this implies, but is stronger than, $\Hh=0$. For example, any simply-connected three-dimensional submanifold $S \subset \Omega$ with $\omega \parallel \partial S$ has a gauge-invariant helicity ($S$ must also transform under any diffeomorphism of $\Omega$). Then it is quite possible to have $\Omega = S_1 \cup S_2$, $\Hh = \Hh_1+\Hh_2=0$ and $\Hh_1 \neq 0$ in which case \eqref{eq:int} cannot be satisfied. We also note that if the integral curves of $\omega$ are closed then \eqref{eq:int} does not imply that the pairwise linking numbers of the vortex lines vanish, for example consider a vorticity field supported on a set of $N$ solid tori, $\gamma_i$ each carrying flux $\Phi_i$, with the vortex lines within each torus having zero mutual linking (zero self-linking term~\cite{moffatt92}). In such a configuration there are $N$ independent gauge invariant measurements of helicity corresponding to integrals of $U \cdot \omega$ over a tubular neighbourhood of each filament. Let $\mathcal{H}_i$ be the helicity corresponding to $\gamma_i$, then \eqref{eq:int} is equivalent to the requirement that all the $\mathcal{H}_i$ vanish. 
\begin{equation}
\mathcal{H}_i = \Phi_i \sum_{j} \Phi_j \textrm{Lk}(i,j) =0,
\end{equation} 
which does not imply $\textrm{Lk}(i,j)=0$ (as an example take all fluxes equal with the tori forming link $8^4_3$ in the Rolfsen table~\cite{rolfsen}).

The case we consider is of an integrable vorticity field, \eqref{eq:int} implies the kernel of $\theta$ defines a (singular) foliation $\mathcal{F}$ of $\Omega$ with $\omega$ tangent to the leaves. If $\theta$ is unique, $\mathcal{F}$ is defined purely in terms of the vorticity field. In Ref.~\cite{enciso16} the integrable case is discussed and diffeomorphisms that act by transforming the pair $(\mathcal{F}, \omega)$ are held distinct from those that act by transforming just the vorticity field. Because $\mathcal{F}$ is defined in terms of the vorticity, this issue is avoided.

For now we will assume that $\theta \neq 0$ and that $\theta$ is the unique 1-form satisfying $d \theta = \omega$ and $\theta \wedge \omega = 0$, then there is a 1-form $\eta$ satisfying,
\begin{equation}
d \theta = \theta \wedge \eta.
\end{equation}
Consider the integral
\begin{equation}
\G = \int_\Omega \eta \wedge d \eta.
\end{equation}
The 1-form $\eta$ is defined up to addition of $f \theta$,  $f$ a function. Under such a transformation we have
\begin{equation}
\G = \int_\Omega \eta \wedge d \eta + \int_\Omega  df \wedge d \theta,
\end{equation}
the second term gives a boundary contribution which vanishes, $\omega$ is tangent to $\partial \Omega$ so $d \theta |_{\partial \Omega}=0$ and $\G$ does not depend on the choice of $\eta$. By construction $\G$ is invariant under any volume-preserving diffeomorphism and hence the fluid flow. 

$\G$ is essentially the Godbillon-Vey invariant~\cite{gv,candel} of the foliation defined by $\theta$. In particular, if $\Omega$ is closed, then $\G$ is exactly the Godbillon-Vey invariant. If $\Omega$ has a boundary then $\G$ is defined for the pair ($\theta, \omega$) whereas the Godbillon-Vey invariant may not be. For a foliation, one requires invariance under the transformation $\theta \to h \theta$ for a non-zero function $h$. This leads to a gauge transformation $\eta \to \eta - d \log h$, and hence to a boundary term 
\begin{equation}
\int_{\partial \Omega} -\log h\; d \eta,
\end{equation}
which vanishes in general only if $\partial \Omega$ is a leaf of the foliation (it is easily verified that $\theta \wedge d \eta =0$), and for an arbitrary integrable vorticity field this is not the case.

\subsection{Non-uniqueness of $\theta$}

So far we have assumed that there is a unique $\theta$ satisfying \eqref{eq:int}. $\theta$ is a potential 1-form for the vorticity and all gauge transformations of $\theta$ are physically equivalent. It follows that if $\theta$ satisfying \eqref{eq:int} is non-unique, then we would like $GV$ to not depend on the choice, so that $GV$ is a well-defined invariant of vorticity. The goal of this section is to show that, restricting to non-zero choices of $\theta$, non-uniqueness imposes sufficient structure on $\omega$ to force $\G=0$.

Suppose the choice of $\theta$ satisfying \eqref{eq:int} is not unique. Then the difference between any two choices is a closed 1-form $\beta$ satisfying $\beta\wedge d \theta=0$, and there is a 1-parameter family of 1-forms, $\theta + c \beta$, $c$ constant, satisfying \eqref{eq:int}. If $\theta \neq 0$ throughout the domain, then there is an open set $ (a,b) \subset \mathbb{R}$, $a<0<b$, such that $\theta + c \beta \neq 0$ for $c \in (a,b)$, this allows us to only consider non-zero $\theta$. Zeros in $\theta$ can be accommodated for, and we will deal with them in part below, but for simplicity we will exclude the possibility for now. 

The 1-form $\beta$ satisfies 
\begin{equation}
\theta \wedge \beta = g d \theta.
\label{eq:gg}
\end{equation}
If the right-hand side of \eqref{eq:gg} vanishes identically, then $h \theta =  \beta$ for some function $h$. $h$ is a first integral of $\omega$, so that $dh \wedge d \theta =0$ and we may take $\eta = d \log h$ on the complement of the zero set of $h$, which we take to be open and dense on $\Omega$. Then let $T_\epsilon = \{x \in \Omega : \; |h(x)|>\epsilon \}$. Then consider the integral
\begin{equation}
\G_\epsilon = \int_{T_\epsilon} \eta \wedge d \eta = 0,
\end{equation}
and as $\epsilon \to 0$, $\G_\epsilon \to \G$ and we find $\G=0$.

Now we consider the case where the right-hand side of \eqref{eq:gg} does not vanish identically. The function $g$ is defined only on the complement of the zero set of vorticity, which we denote $Z_\omega = \{ x \in \Omega :\, \omega = 0 \}$, and we take the complement $Z_\omega^c$ open and dense. We also define $Z_g$ to be the zero set of $G$, which we also take to have open and dense complement. Now consider the 1-forms $\beta / g$, and $\theta^\prime = f \theta$ on $Z_\omega^c \cap Z_g^c$, $f$ an arbitrary function. These satisfy
\begin{equation}
\frac{\beta}{g} \wedge d \theta^\prime = d f \wedge d \theta,
\label{eq:what}
\end{equation}
$f$ is arbitrary and the right-hand side is smooth on all of $\Omega$. In particular, we may choose $f$ so that at some point $p \in Z_\omega \cup Z_g$, $d \theta^\prime $ is an arbitrary 2-form satisfying $\theta \wedge d \theta^\prime =0$. This implies that $\beta / g = \xi + k \theta$ where $\xi$ is smooth on $\Omega$ and $k$ is potentially singular (on the zero sets of both $g$ and $\omega$). 

We choose $ \eta = \xi $, then $\eta \wedge d \eta$ is smooth on the entire domain, but on $Z_\omega^c \cap Z_g^c$ we may write $\eta \wedge d \eta = dk \wedge d \theta$, which must then extend smoothly on to $\Omega$. Now let $k_\epsilon$ be a smooth approximation to $k$. We can integrate $dk_\epsilon \wedge d \theta$ over $\Omega$ to give $GV_\epsilon=0$. As $\epsilon \to 0$, $GV_\epsilon \to GV$, so we conclude $GV = 0$. 

Finally one could consider the mixed case, where the zeros of $g d \theta$ have some interior, $S$. If $S$ is sufficiently well-behaved, then $\omega$ must be tangent to the boundary of $S$ and one can treat the two regions, $S$ and $S^c$, separately using the above methods.

As a consequence of this, we find that the value of $\G$ depends purely on the vorticity field, rather than the 1-form $\theta$, if $\theta$ is not unique then $\G=0$. Finally, note that this result also forces $\G=0$ if $\omega$ has a first integral.

\section{$\G$ and Dynamics}

\subsection{Global Obstruction to Steady Flow}

$\G$ has natural relation to the dynamics of the flow. First suppose the flow is steady, and that there is a choice of non-zero $\theta$ satisfying \eqref{eq:int}, then following Arnold~\cite{arnold99}, the flow must be one of three types:
\begin{enumerate}
\item $U \times \omega  = X$, where $X$ is curl-free.
\item Non-constant Beltrami field, $\omega = \alpha U$, with $\alpha$ a non-constant function.
\item Constant Beltrami field, $\omega = \lambda U$ with $\lambda$ constant. \end{enumerate}

In the first two cases $\theta$ is not unique, the dual to $X$ is a closed 1-form $\beta$ satisfying $\beta \wedge d \theta =0$, and the function $\alpha$ is a first integral of $\omega$, so $\alpha \wedge d \theta =0$. In both these cases, $\G=0$ by arguments in the previous section. In the third case, helicity is non-zero if $\lambda \neq 0$, so $\G$ cannot be defined. If $\lambda = 0$, then the flow is irrotational, and $\G=0$ trivially.

Invariant under volume-preserving diffeomorphisms, $\G \neq 0$ is then a global obstruction to the existence of a metric making $\omega$ a steady solution of the Euler equations, under the assymptions given. This can be compared with results on the topology of the vorticity scalar in two dimensions obstructing steady flow~\cite{ginzburg94}; results on the `Eulerisability' of velocity fields~\cite{ps19,cieliebak14}, where it is shown that such velocity fields cannot contain Reeb components; as well as results making use of contact topology, showing that a vector field without a closed orbit cannot be a solution to the Euler equations (for analytic vector fields on $S^3$)~\cite{etnyre00}. 


\subsection{Local Obstruction to Steady Flow}

We now give an alternate version of the connection between $\G$ and dynamics. For simplicity we assume that $\Omega$ is a subset of $\mathbb{R}^3$ with the standard metric and total volume $V$. Then let the vector field $H$ be dual to $\eta$ (in coordinates $H^i g_{ij} = \eta_j$, for $g$ the metric), giving
\begin{equation}
\G= \int_\Omega H \cdot \nabla \times H\; dV.
\end{equation}
For $h$ we make the choice
\begin{equation}
H =\frac{1}{U \cdot A} \omega \times U,
\label{eq:hchoice}
\end{equation}
where $A$ is the vector field dual to $\theta$ (in coordinates $A^i g_{ij} = \theta_j$) and note that it is well-defined as per Section~\ref{section:2}, provided $U \cdot A \neq 0$, if $U \cdot A = 0$, one can modify $H$ in the neighbourhood of such a zero. Assuming $U \cdot A \neq 0$ throughout and noting that the vorticity equation gives $\partial_t \omega = - \nabla \times  (\omega \times U)$ we find
\begin{equation}
H \cdot \nabla \times H = -\frac{1}{(U \cdot A)^2} ( \omega \times U) \cdot \partial_t \omega.
\end{equation}
With this choice of $H$, $\partial_t \omega$ is non-zero wherever $H \cdot \nabla \times H$ does not vanish, the density of $\G$ gives a local obstruction to the flow being steady and implies the global result. This perspective also translates to arbitrary vector fields $V$ transporting the vorticity. One can replace $U$ in \eqref{eq:hchoice} with any volume preserving vector field $V$, whereupon the density of $\G$ is related to the time derivative of $\omega$ under the flow of $V$.

\subsection{Bounds on Vorticity Rate-of-Change}

Helicity is a dimensionful quantity, if $U$ has dimensions $[X]$, then helicity has dimensions $[X]^2[\textrm{Length}]^2$. Under a rescaling $\omega \to c\; \omega$, helicity transforms as $H \to c^2 H$ (along with orientation reversal, this is sufficient to establish a bijection between the sets of vector fields with fixed non-zero helicity) and it is well-known that helicity bounds the enstrophy $\int_\Omega \omega^2 \;dV$~\cite{arnold74}. Conversely, $\G$ is dimensionless (note that $\eta$ always has dimensions of $\textrm{Length}^{-1}$) so can only bound ratios of physical quantities. First note that 
\begin{equation}
\int_\Omega U \cdot A \;dV  = \int_\Omega U^2 dV = 2 E,
\end{equation}
where $E$ is the kinetic energy of the flow, so we can write $ U \cdot A = 2E/V + \delta$, expanding we find
\begin{equation}
GV = \frac{V^2}{4 E^2} \int_\Omega dV (\omega \times U) \cdot (\partial_t \omega) \left ( 1 - \delta \frac{ V}{E} +  O(\delta^2)  \right ).
\end{equation}
Neglecting terms of order $\delta$ and above (constant energy density) we find
\begin{equation}
GV \approx \frac{V^2}{4 E^2} \int_\Omega  (\omega \times U) \cdot (\partial_t \omega) dV.\end{equation}
Then using Cauchy-Schwarz and the Poincar\'{e} inequality we find
\begin{equation}
|GV| \lessapprox \frac{L^7}{4 E^2} \int_\Omega (\partial_t \omega)^2 dV,
\end{equation}
where $L^7 = V^2 /\sqrt{\lambda}$ and $\lambda$ is the minimal Laplacian eigenvalue, so that $GV$ is related to the ratio of the rate of change of the vorticity to the (squared) kinetic energy.


\section{Zeros of $\theta$}

So far we have required $\theta \neq 0$. Zeros in $\theta$ can be accommodated, and we will discuss some aspects here. Suppose $\theta$ has zero set $\mathcal{Z}$, so that $\theta$ defines a singular foliation of $\Omega$. We will deal with two particular types of zeros. The first, Morse-type, are isolated zeros of vorticity where $\theta = f dg$, and $dg$ has a critical point. The second type correspond to particular closed orbits of vorticity, around which $\theta$ has the form given in \eqref{eq:kupka}, and one can think of as local vortex filament structures within the flow.

While zeros in general can be dealt with through the use of Haefliger structures~\cite{ghys89}, we will deal with the differential forms directly. Structurally stable, zeros of integrable 1-forms have a rich structure~\cite{camacho78,camacho82}. For example, the 1-form
\begin{equation}
\theta = \alpha y z d x+ \beta x z dy+\gamma x y dz,
\label{eq:lab}
\end{equation}
with $\alpha \neq \beta \neq \gamma$ is homogeneous of degree 2, satisfies $\theta \wedge d\theta=0$ and has a $C^2$-stable zero at the origin. A related example has also been used to demonstrate the local non-existence of a Clebsch representation for velocity fields in the vicinity of vorticity zeros~\cite{graham00}. In general, in the smooth category, there are structurally stable examples such as \eqref{eq:lab} of arbitrarily high degree. Here we restrict our discussion to zeros of $\theta$ that are $C^1$ structurally stable, and note that an extension to higher-order zeros should be possible. $C^1$ structurally stable zeros come in two varieties~\cite{kupka64}. The first are points where $\theta$ has the local form $f dg$, with $f\neq0$ and $g$ having a Morse critical point. In a neighbourhood of such a point, one can write $\eta = -d \log f$ which is non-singular, so that $\eta$ can be defined.

The second type of zero has $\theta=0$ and $d \theta \neq 0$, in this case the Kupka phenomenon occurs~\cite{kupka64,camacho78,camacho82}, and in a neighbourhood of such a zero there is a local system of coordinates such that 
\begin{equation}
\theta = a(x,y) dx+b(x,y) dy,
\label{eq:kupka}
\end{equation}
where $a$ and $b$ have a simple zero at $x=y=0$. In this case the zero becomes a line $L$. $\omega$ is tangent to $L$, which therefore cannot intersect $\partial \Omega$ transversely, making $L$ a distinguished closed orbit. On such a Kupka line $\eta$ cannot be defined, but $\G$ can still be computed. From \eqref{eq:kupka} we can define a neighbourhood,  $N_{L}$, such that $d \theta|_{\partial N_L} = 0$. The torus $\partial N_{L}$ separates the domain into two regions, as the vorticity is tangent to $\partial N_{L}$ both the exterior and interior (containing the Kupka line) have a well-defined $\G$, so we may write $\G = \G_e+\G_i$, for exterior and interior respectively. It is then simply a matter of checking (which can be done by a simple calculation), that $\eta$ may be defined on the interior of $\partial N_L$ such that $\eta \wedge d \eta$ is identically zero on the complement of $L$, and so we may identify $\G_i=0$.

It is important to note that if we allow more complex zeros in $\theta$, then the analysis becomes more delicate. For example, suppose we allow \eqref{eq:lab}. Then $\eta$ cannot be defined on the zero. In this case $\G$, as defined above, is invariant under $\eta \to \eta+ f \theta$ only if $f  \sim 1/\|x\|^{a}$, as $x \to 0$, where $a < 3$.


\section{Local Structure}

\subsection{Local Conservation Law}
Like helicity, $\G$ is a global property of the fluid and cannot be measured locally. For example, in a neighbourhood of any point where $\omega \neq 0$ we have a Clebsch representation~\cite{graham00},
\begin{equation}
U = f \nabla g+ \nabla \phi.
\label{eq:cl}
\end{equation}
Consequently, one may choose a vector potential for $\omega$ as well as a particular $\eta$ such that both the helicity density and the Godbillon-Vey density vanish on the neighbourhood. Despite this, there are several local analyses that are informative. First we will derive a local conservation law for $\G$, which obtains validity by making a global choice for $\eta$. 

Suppose $\Omega \subset \mathbb{R}^3$ with the Euclidean metric. Then in a neighbourhood about any point with $\omega \neq 0$, we may use \eqref{eq:cl} and write $\theta = f d g$. Note that in the Clebsch representation the functions $f$ and $g$ are both transported by the flow~\cite{vez} so that
\begin{equation}
(\partial_t + \mathcal{L}_U) f = (\partial_t + \mathcal{L}_U) g = 0
\end{equation}
then by properties of the Lie derivative we find
\begin{equation}
(\partial_t + \mathcal{L}_U) \theta = \left ((\partial_t + \mathcal{L}_U) f \right ) d g + f d \left ( (\partial_t + \mathcal{L}_U) g  \right ) = 0
\label{eq:thtr}
\end{equation}
since the point around which we used the Clebsch expansion was arbitrary, we find that $\theta$ is transported by the flow. We note that this can also follows as $\theta$ is defined as a vector potential satisfying $\theta \wedge d \theta = 0$. Since $d \theta$ is transported by the flow via the vorticity equation ($(\partial_t + \mathcal{L}_U)  d \theta = 0$), we must have (locally) $(\partial_t + \mathcal{L}_U)   \theta = d c$, for some function $c$. But $\theta$ is defined by the requirement that $\theta \wedge d \theta = 0$ which must be constant in time, hence $d c  =0$, so that \eqref{eq:thtr} holds.

The vorticity equation reads $(\partial_t+\mathcal{L}_U)d \theta=0$, using $d\theta = \theta \wedge \eta$ and the result that $(\partial_t+\mathcal{L}_U)\theta=0$ we have
\begin{equation}
(\partial_t + \mathcal{L}_U) \eta = k \theta
\end{equation}
for some function $k$. It follows that
\begin{equation}
(\partial_t + \mathcal{L}_U) \eta \wedge d \eta  = d(k d \theta).
\end{equation}
In vector calculus notation we have
\begin{equation}
(\partial_t + U \cdot \nabla) (H \cdot \nabla \times H) = \nabla \cdot (k \omega)
\end{equation}
which gives an explicit local conservation law, and we find that $\G$ is carried by vorticity. One must choose $H$ to determine $k$. For example, taking $H$ as in \eqref{eq:hchoice} gives the result
\begin{equation}
k =  H^2 + \frac{1}{U \cdot A}  H \cdot \nabla(P+U^2/2),
\end{equation} 
or
\begin{equation}
k =  \frac{1}{(U \cdot A)^2}\left ( \omega^2 U^2 - (\omega \cdot U)^2+ (\omega \times U) \cdot \nabla(P+U^2/2) \right),
\end{equation} 
where $P$ is the pressure. The analysis by Tur and Yanovsky~\cite{tur93} yields an alternate form for $k$ when the choice $H \cdot A=0$ is made.

\subsection{Interpretation as Helical Compression}

Due to the form of $\G$ it is tempting to attempt to interpret it as the linking of the integrals curves of the vector field $X$ associated to the 2-form $d \eta$ as $d \eta = \iota_X \mu$. This is not natural for two reasons. Firstly, $d \eta$ is not necessarily parallel to $\partial \Omega$, so integral curves of $X$ may end. Secondly, $d \eta$ is not invariant under the transformation $\eta \to \eta + f \theta$, so that the integral curves of $X$ change, and may reconnect locally. Instead, there is a local geometric interpretation of $\G$ as helical compression of vorticity. Vorticity compression is the core nonlinearity of the Euler equations, and given the strong relationship between $\G$ and dynamics, we speculate that flows with $\G \neq 0$ will be particularly interesting from a dynamical perspective. The helical compression interpretation, essentially Thurston's `helical wobble' for foliations~\cite{webb19,thurston72}, can be seen as follows. Suppose $\mathcal{F}$ is the foliation induced by $\omega$, and $N$ is a unit vector normal to it. Then the vector field $h_N = (N \cdot \nabla) N$ is a choice of $h$ for $\nabla \times N$, and is hence equal to a choice of $h$ for $\omega$ up to a gauge transformation. 

\begin{figure}
\begin{center}
\includegraphics{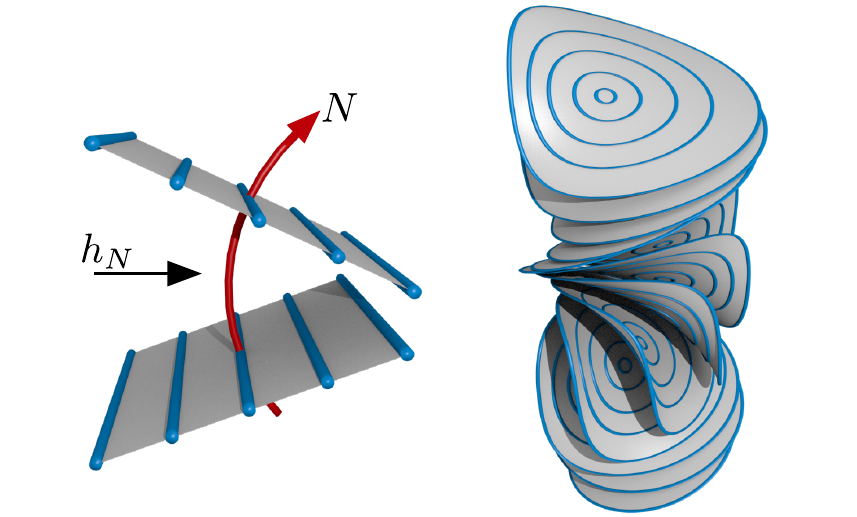}
\end{center}
\caption{Local geometric interpretation of $GV$ as helical vortex compression. Left: $N$ (red) is the normalised field to the foliation $\mathcal{F}$ to which the vorticity (blue) is tangent. The vector $h_N$ points in the direction of local compression of the surfaces of $\mathcal{F}$. Right: helical compression of $\mathcal{F}$. In this case the vector $h_N$ rotates about the vertical as one moves upwards, hence $h_N \cdot \nabla \times h_N \neq 0$.}
\label{fig:helical_fig}
\end{figure}

$h_N$ measures the local compression of the leaves of $\mathcal{F}$, as shown in Figure~\ref{fig:helical_fig}(left), its magnitude is the curvature of the integral curves of the normal vectors to $\mathcal{F}$, and it points along the direction of local expansion. Twisting of $h_N$ as one moves transverse to the leaves of $\mathcal{F}$ (Figure~\ref{fig:helical_fig}(right)) corresponds to helical compression of the vortex lines and also $h_N \cdot \nabla \times h_N \neq 0$. Following Reinhardt and Wood~\cite{reinhardt73}, one can give a local pointwise expression for $h_N \cdot  \nabla \times h_N$ as $\kappa^2 (\tau - B \cdot \nabla N \cdot  Z)$, where $\kappa$ and $\tau$ are the curvature and torsion of the integral curves of $N$ and $B$ and $Z$ are the binormal and normal fields for the integral curves of $N$ in the Frenet-Serret frame. Because it can be interpreted as a combination of twisting and compression, we are tempted to refer to $\G$ as the `wring' of the vorticity.

\section{Global Structure}

For $\G \neq 0$ one must construct a vorticity field with global properties, much like the helicity requires the global linking of vortex lines. All known examples of vorticity fields with $\G \neq 0$ have a certain degree of complexity, and do not permit a simple expression. Here we will adapt Thurston's construction~\cite{thurston72} to give an example of a vorticity field where $\G$ is determined by certain Kupka lines -- closed orbits of $\omega$.

\begin{figure}
\begin{center}
\includegraphics[width=0.5\textwidth]{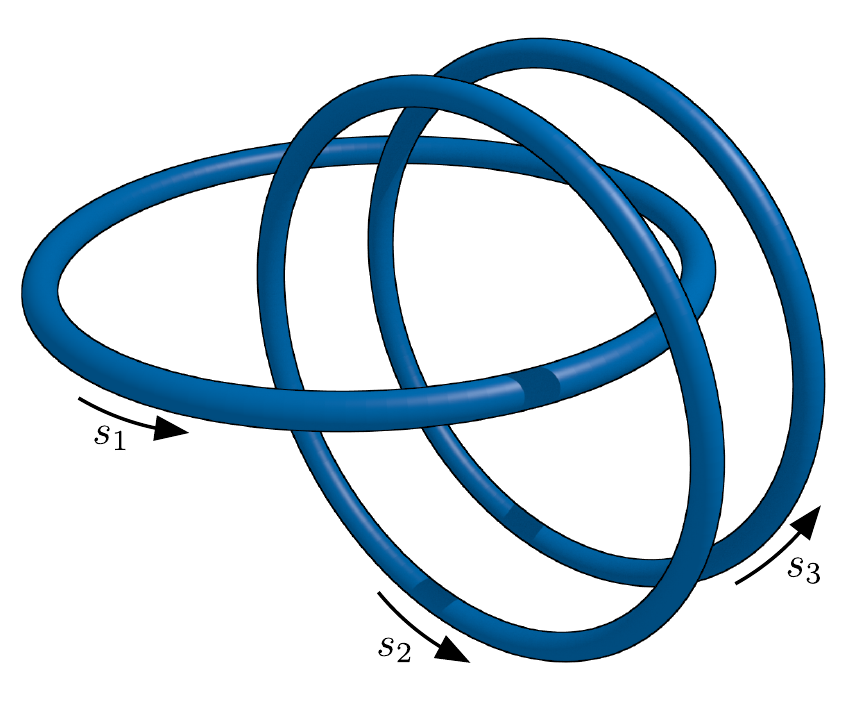}
\end{center}
\caption{The link $L$ with 3 components in $S^3$. In the example given, each component of $L$ is a closed orbit of $\omega$. The local structure of $\theta$ in a neighbourhood of $L$ defines a slope for each component. The value of $\G$ is determined by the deviation of these slopes from the value they would have in the case when $\omega$ is singular, with each $L$ a flux line.}
\label{fig:link}
\end{figure}

First consider the link, $L$, in $S^3$ with $N$ components each denoted $L_i$, $i \in [1,N]$, shown in Figure~\ref{fig:link} for $N=3$. Thurston's construction uses as its base the horocycle foliation of the unit tangent bundle of the hyperbolic plane\footnote{In the Poincar\'{e} disk model, this is defined by the 1-form $\alpha = 2 (y - \sin \theta) dx + 2 (\cos \theta -x) d y + ( x^2 + y^2-1) d \theta$; where $\theta$ is the fiber coordinate. The leaves of the foliation are helicoids with axis on the boundary.} and creates a foliation on the complement of this link, defined by a 1-form $\theta$. The leaves of the foliation are transverse to the boundary and induce a slope $s_i$ on each link component (measured relative to the oriented longitudes as in Figure~\ref{fig:link}). In Thurston's construction, the boundary components are spun~\cite{candel} onto tori and filled with Reeb components. Equivalently, in our case, we can simply allow them to become zeros of $\theta$, where they become Kupka lines. Since the neighbourhood of each Kupka line has a well defined value of $\G =0$, the value of $\G$ for the total vorticity field must be equal to the Godbillon-Vey invariant of the foliation, and is given by
\begin{equation}
\G = 4 \pi^2 \left ( N-2 -\left( \frac{1}{s_1} +\sum_{i=2}^N s_i \right ) \right).
\label{eq:thurston}
\end{equation}
To understand the meaning of \eqref{eq:thurston}, we construct a $\G=0$ vorticity field with the same set of closed vortex lines. First we define a closed 1-form $A$ on $S^3 \setminus L$. As $H^1(S^3 \setminus L;\mathbb{R}) \cong \mathbb{R}^{N}$, this form is defined by a set of $N$ fluxes $\phi_i$, and leads to a measured (singular) foliation of $S^3 \setminus L$. We can then multiply $A$ by a function $f$, zero on $L$, so that $A$ can be extended to a 1-form on $S^3$. The resulting singular foliation of $S^3$ has $\G=0$ and $N$ Kupka lines, with slopes $s_1 = -\phi_1/\sum_{i \neq 1} \phi_i$ on $L_1$ and $s_i = -\phi_i/\phi_1$ on $L_i$, $i \neq 1$. Then note that $\sum_{i =2}^N s_i +1/s_1=0$. Comparing to \eqref{eq:thurston}, we see that $\G$ measures, in part, the failure of the Kupka line slopes to `commute' in the manner expected for singular vorticity field with flux lines on $L$.

The construction has at its core a foliation that is transverse to the fibers of a circle bundle, in which case $\eta\wedge d \eta$ can be integrated over the fibers~\cite{bott78}. Using similar techniques~\cite{brooks79} it should be possible to a range of examples similar examples to the one above where the Kupka lines form any link $L$ whose complement is an $S^1$ bundle over $S^2$. These have been classified~\cite{burde70}, which in principle allows for such a construction to be made.

It is a pleasure to thank J.H.~Hannay for many stimulating conversations, as well as D. Peralta-Salas for his hospitality and insight. I would also like to acknowledge useful discussions with P.J.~Morrisson, G.P.~Alexander, D.~Barkley, M.V.~Berry and H.K.~Moffatt.


\section{Appendix I}

We consider the invariance of helicity
\begin{equation}
H = \int_\Omega \theta \wedge d \theta,
\end{equation}
under an arbitrary transformation, $\theta \to \theta + \beta$, where $\beta$ is a closed 1-form. Then the helicity integral gives a boundary term
\begin{equation}
H \to \int_{\Omega} \theta \wedge d \theta + \int_{\partial \Omega} \theta \wedge \beta,
\end{equation}
which in general does not vanish. By construction, $d \theta |_{\partial \Omega}=0$, hence $\theta|_{\partial \Omega}$ is a closed 1-form on $\partial \Omega$ and defines a de Rahm cohomology class $[\theta] \in H^1(\partial \Omega ; \mathbb{R})$. Restriction to the boundary defines a map $r:H^1(\Omega ;\mathbb{R}) \to H^1(\partial \Omega ; \mathbb{R})$. $d \theta$ is fluxless if $[\theta] =0 \in   H^1(\partial \Omega ; \mathbb{R}) / \textrm{Im}(r) $. This condition is equivalent to the statement that
\begin{equation}
\int_{S} d \theta =0,
\end{equation}
for any surface $S \subset \Omega$, with $\partial S \subset \partial \Omega$.

\section{Appendix II}

We start with the definition
\begin{equation}
H= \frac{1}{U \cdot A} \omega \times U = \beta \omega \times U.
\end{equation}
Then we assert that
\begin{equation}
(\partial_t + \mathcal{L}_U) H = f A,
\end{equation}
where $f$ is to be determined. Using coordinate notation (recall we are in Euclidean space, so that we do not distinguish covariant and contravariant indices), we have
\begin{equation}
f A_i = \partial_t H_i + U_j \partial_j H_i +  U_j\partial_i H_j.
\end{equation}
Now by construction we have $U_i H_i=0$, so this is rewritten as
\begin{equation}
f A_i= \partial_t H_i + U_j \partial_j H_i - H_j\partial_i U_j,
\end{equation}
or
\begin{equation}
f A=\partial_t H - U \times \nabla \times H.
\end{equation}
Expanding we find
\begin{equation}
f A = (\partial_t \beta) \omega \times U + \beta (\partial_t\omega)\times U + \beta \omega \times (\partial_t U) - U \times \big( \nabla \beta \times (\omega \times U) \big) + \beta U \times (\partial_t \omega),
\end{equation}
(recall $\beta= (U \cdot A)^{-1}$) which becomes
\begin{equation}
f A =((\partial_t +U \cdot \nabla)\beta) \omega \times U + \beta \omega \times (\partial_t U).
\end{equation}
Now, using the fact that
\begin{equation}
(\partial_t + U \cdot \nabla)A + (\nabla A) \cdot U =0,
\end{equation}
we find
\begin{equation}
((\partial_t +U \cdot \nabla)\beta) = \beta (H \cdot A + \beta A \cdot \nabla(P+U^2/2)).
\end{equation}
So we get
\begin{equation}
f A = \left ((h \cdot A) + \beta  A \cdot \nabla(P+U^2/2) \right ) H - \beta \omega \times \nabla (P+U^2/2) - \beta \omega \times (\omega \times U).
\end{equation}
Then we find 
\begin{equation}
f A = \left ( H^2 + \beta  H \cdot \nabla(P+U^2/2)  \right) A,
\end{equation}
so we may identify
\begin{equation}
f =  H^2 + \beta  H \cdot \nabla(P+U^2/2) .
\end{equation}

\end{document}